\documentstyle[general_cite,psfig]{mn}
\title{The entropy and energy of intergalactic gas in galaxy
clusters}
\author[E. J. Lloyd-Davies et al.]
       {E. J. Lloyd-Davies,\thanks{E-mail: eld@star.sr.bham.ac.uk} 
        T. J. Ponman and D. B. Cannon\\
        School of Physics and Astronomy, University of
        Birmingham, Edgbaston, Birmingham B15 2TT, UK\\}
\date{Accepted 1999 ??.
      Received 1999 ??;
      in original form 1999 ??}

\pagerange{\pageref{firstpage}--\pageref{lastpage}}
\pubyear{1999}

\begin{document}

\maketitle

\label{firstpage}

\begin{abstract}
  Studies of the X-ray surface brightness profiles of clusters, coupled
  with theoretical considerations, suggest that the breaking of
  self-similarity in the hot gas results from an `entropy floor',
  established by some heating process, which affects the structure of the
  intracluster gas strongly in lower mass systems.  By fitting analytical
  models for the radial variation in gas density and temperature to X-ray
  spectral images from the \emph{ROSAT} PSPC and \emph{ASCA} GIS, we have
  derived gas entropy profiles for 20 galaxy clusters and groups. We show
  that when these profiles are scaled such that they should lie on top of
  one another in the case of self-similarity, the lowest mass systems have
  higher scaled entropy profiles than more massive systems.  This appears
  to be due to a baseline entropy of 70-140~${h_{50}}^{-\frac{1}{3}}$ keV
  cm$^{2}$ depending on the extent to which shocks have been suppressed in
  low mass systems.  The extra entropy may be present in all systems, but
  is detectable only in poor clusters, where it is significant compared to
  the entropy generated by gravitational collapse. This excess entropy
  appears to be distributed uniformly with radius outside the central
  cooling regions.
  
  We determine the energy associated with this entropy floor, by studying
  the net reduction in binding energy of the gas in low mass systems, and
  find that it corresponds to a preheating temperature of $\sim 0.3$~keV.
  Since the relationship between entropy and energy injection depends upon
  gas density, we are able to combine the excesses of 70-140~keV~cm$^{2}$
  and 0.3~keV to derive the typical electron density of the gas into
  which the energy was injected. The resulting value of 1-3$\times
  10^{-4}{h_{50}}^{\frac{1}{2}}$\,cm$^{-3}$, implies that the heating must
  have happened prior to cluster collapse but after a redshift $z\sim$
  7-10.  The energy requirement is well matched to the energy from
  supernova explosions responsible for the metals which now pollute the
  intracluster gas.
\end{abstract}


\section{Introduction}
The hierarchical clustering model for the formation of structure in the
universe predicts that dark matter halos should be scaled versions of each
other\cite{navarro95a}. While some energy transfer between dark matter and
gas is possible through gravitational interaction and shock heating,
simulations suggest that the gas and dark matter halos will be almost
self-similar in the absence of additional heating or cooling processes
\cite{eke98a}. Comparison of the structure of real galaxy systems with this
predicted self-similarity provides an excellent probe of extra physical
processes that may be taking place in galaxy clusters and groups.

It has been suggested that specific energy in cluster cores is
higher than expected from gravitational collapse and that this may be due
to energy injected by supernova-driven protogalactic winds
\cite{white91a,david91a}. \scite{david96a} studied the entropy in a small
sample of galaxy systems and suggested that the entropy in their cores had
been flattened due to energy injection.  \scite{ponman98a} have recently
shown that the surface brightness profiles of clusters and groups do not
follow the predicted self-similar scaling.  Surface brightness profiles of
galaxy groups are observed to be significantly flatter than those of
clusters, indicating differences in the gas distribution.

\begin{table*}
\begin{minipage}{154mm}
\caption{Some important properties for the sample of 20 galaxy clusters and groups. Systems are
  listed in order of increasing temperature.}
\label{tab:sample}
\begin{tabular}{lccccccc}
\hline
\hline
Cluster/Group&R.A.(J2000)&Dec.(J2000)&z&$N_{H}$($10^{20}$ cm$^{-2}$)&$T$(keV)&$Z$(solar)&Data\\
\hline
HCG 68&208.420&40.319&0.0080&0.90&0.54&0.43&PSPC\\
HCG 97&356.845&-2.169&0.0218&3.29&0.87&0.12&PSPC\\
HCG 62&193.284&-9.224&0.0137&3.00&0.96&0.15&PSPC\\
NGC 5044 Group&198.595&-16.534&0.0082&5.00&0.98&0.27&PSPC\\
RX J0123.6+3315&20.921&33.261&0.0164&5.0&1.26&0.33&PSPC\\
Abell 262&28.191&36.157&0.0163&5.4&1.36&0.27&PSPC\\
IV Zw 038&16.868&32.462&0.0170&5.3&1.53&0.40&PSPC\\
Abell 400&44.412&6.006&0.0238&9.1&2.31&0.31&PSPC\\
Abell 1060&159.169&-27.521&0.0124&5.01&3.24&0.27&PSPC+GIS\\
MKW 3s&230.507&7.699&0.0453&3.1&3.68&0.30&PSPC\\
AWM 7&43.634&41.586&0.0173&9.19&3.75&0.33&PSPC\\
Abell 780&139.528&-12.099&0.0565&4.7&3.8&0.23&PSPC+GIS\\
Abell 2199&247.165&39.550&0.0299&0.87&4.10&0.30&PSPC\\
Abell 496&68.397&-13.246&0.0331&4.41&4.13&0.31&PSPC+GIS\\
Abell 1795&207.218&26.598&0.0622&1.16&5.88&0.26&PSPC\\
Abell 2218&248.970&66.214&0.1710&3.34&6.7&0.20&PSPC+GIS\\
Abell 478&63.359&10.466&0.0882&13.6&7.1&0.21&PSPC+GIS\\
Abell 665&127.739&65.854&0.1818&4.21&8.0&0.28&PSPC+GIS\\
Abell 1689&197.873&-1.336&0.1840&1.9&9.0&0.26&PSPC+GIS\\
Abell 2163&243.956&-6.150&0.2080&11.0&13.83&0.19&PSPC+GIS\\ 
\hline
\end{tabular}
Notes: Positions, hydrogen columns and redshifts are taken from
\scite{ebeling96a}, \scite{ebeling98a}, \scite{ponman96a} and
\scite{helsdon99a}. Emission weighted temperatures and metallicities for
these systems are taken from \scite{markevitch96a}, \scite{mushotzky97a},
\scite{markevitch98a}, \scite{mchardy90a}, \scite{fukazawa98a},
\scite{butcher91a}, \scite{mushotzky97b}, \scite{helsdon99a},
\scite{david96a}, \scite{david94a} and \scite{ponman96a}.
\end{minipage}
\end{table*}

In order to explore this effect further, it is necessary to study the
properties of the gas in these systems in greater detail. A particularly
interesting property of the gas for this purpose is its entropy, as this
will be conserved during adiabatic collapse of the gas into a galaxy
system, but is likely to be altered by any other physical processes. For
instance preheating of the gas before it falls into the cluster, energy
injection from galaxy winds and radiative cooling of the gas in dense
cluster cores will all perturb the entropy profiles of clusters from the
self-similar model. Analysis of the entropy profiles of virialized systems
of different masses should therefore allow the magnitude of such effects to
be studied, constraining the possible processes responsible. A key question
to answer in this regard is how much energy is involved in any departures
from self-similarity of the entropy profiles. The study of
\scite{ponman98a} was not able to address this issue in detail, since the
gas was assumed to be isothermal.  Here we combine \emph{ROSAT} PSPC and
\emph{ASCA} GIS data to constrain temperature profiles, allowing a more
detailed study of entropy and energy distributions in the intergalactic
medium (IGM).

Energy loss from the gas due to cooling flows in the centres of clusters
and groups will actually lead to an increase in the gas entropy outside the
cooling region \cite{knight97a}. This is because as gas cools out at the
centre of the system, gas from a larger radius, which has higher entropy,
flows in adiabatically to replace it. However this effect will not be very
large unless a significant fraction of the gas in the system cools out,
which is not feasible within a Hubble time, even for systems with
exceptionally large cooling flows. 

Energy injection into the gas will also raise the entropy profiles of
systems. This energy injection could occur either before or after the
systems' collapse, but more energy is needed to get the same change in
entropy when the gas is more dense \cite{ponman98a}. There are several
possible processes that might have injected energy into the intracluster
medium: radiation from quasars, early population III stars, or energetic
winds associated with galaxy formation. 

There may also be transient effects on the entropy profiles of
systems due to recent mergers. Hydrodynamical simulations suggest that the
entropy profiles of systems are flattened and their central entropy raised
during a merger, and this will last until the system settles back into
equilibrium \cite{metzler94a}. In order to look for the effects of extra
physical processes, it is advantageous to study a set of systems with a
large range in system mass, as these processes will break the expected
self-similar scaling relations. In the present paper we examine the
properties of the intracluster gas in systems with mean temperatures
ranging over a factor of 25, corresponding to virial masses varying by over
two orders of magnitude.

\section{Sample}
The sample selected for this study consisted of 20 galaxy systems ranging
from poor groups to rich clusters, with high quality \emph{ROSAT} PSPC and
in some cases \emph{ASCA} GIS data.  Basic properties of these systems are
listed in Table \ref{tab:sample}. The sample was chosen to cover a wide
range of system masses but is not a `complete' or statistically
representative sample of the galaxy cluster/group population. It is
necessary that the systems be fairly relaxed and spherical in order for the
assumption of spherical symmetry used in the analysis to be reasonable, and
they were selected with this in mind, although it will be seen later that
some of the systems are not as relaxed as we had hoped. In general, our
sample should be representative of the subset of galaxy systems which is
fairly relaxed and X-ray bright. Galaxy systems which are not virialized,
or those currently undergoing complex mergers, would be expected to have
systematically different properties. Our sample spans the population range
from small groups to rich clusters, covering a range in emission weighted
gas temperature from 0.5 to 14 keV. It is therefore well-suited to
investigating the scale dependence in cluster properties.

\section{Data reduction}
In general \emph{ASCA} GIS data were used only where \emph{ROSAT} PSPC data
were insufficient to constrain the models. This was generally the case for
systems with temperatures greater than 4 keV but the cutoff can be somewhat
higher for high quality \emph{ROSAT} PSPC data (i.e. Abell 1795 and Abell
2199).  In the cases where it is possible to access the consistency of
results from \emph{ROSAT} PSPC and \emph{ASCA} GIS data it appears that
they are in reasonable agreement.  The results of fits to \emph{ROSAT} PSPC
data and joint fits to \emph{ROSAT} PSPC and \emph{ASCA} GIS data for Abell
1060 are quite similar. The temperature profiles derived from \emph{ROSAT}
PSPC data for Abell 1795 and Abell 2199 are also consistent with the
emission weighted temperature obtained by previous authors from \emph{ACSA}
data.

A similar reduction process was applied to the \emph{ROSAT} and \emph{ASCA}
data for each system. For the \emph{ROSAT} PSPC, the data were screened to
remove periods of high particle background, where the master veto rate was
above 170 counts s$^{-1}$. The background was calculated from an annulus
typically between 0.6-0.7$^{\circ}$ off-axis.  This annulus was moved to
larger radii for clusters of large spatial extent, to avoid cluster
emission contaminating the background. Point sources of significance
greater than 4$\sigma$, together with the PSPC support spokes, were removed
and the background in the annulus was extrapolated across the detector
using the energy dependent vignetting function.

For the \emph{ASCA} GIS, the data were screened to to remove periods of
high particle background. The following parameters were used to select good
data; cut-off rigidity ($COR$) $>$ 6 GeV c$^{-1}$; radiation belt monitor
count rate $<$ 100; GIS monitor count rate `H02' $<$ 45.0 and $<$ 0.45 x
$COR^{2}$ - 13 $\times$ $COR$ + 125. Data were also excluded where the
satellite passed through the South Atlantic Anomaly and where the elevation
angle above the Earth's limb was $\leq$ 7.5$^{\circ}$. The background was
taken from the sum of a number of `blank sky' fields screened in the same
way as the source data and scaled to have the same exposure time as the
observation of the source.

In order to carry out our cluster modelling analysis, spectral image cubes
were sorted from the raw data. The \emph{ROSAT} PSPC cubes had 11 energy
bins covering PHA channel 11 to 230, and spatial bins $25''$ in size. The
\emph{ASCA} GIS cubes had 24 energy bins spanning PHA channel 120 to 839,
and spatial bins $1.96'$ in size.  Only data from within the PSPC support
ring were used. For all systems this encompassed the great majority of the
detectable \emph{ROSAT} flux. PSPC radial surface brightness profiles were
used to set the extraction radius in each case to restrict data to the
region where diffuse emission is apparent above the noise. \emph{ASCA} data
were extracted from a regions similar in size to the corresponding PSPC
dataset.  Point sources were removed from the PSPC cubes. In the case of
\emph{ASCA}, the poor PSF makes this infeasible, however none of our
targets includes bright hard sources which might seriously affect our GIS
analysis.

The data cubes were background subtracted and then normalized to 
count~s$^{-1}$.  The cubes were not corrected for vignetting as this would
invalidate the Poisson statistics assumed in our subsequent analysis.
Instead the vignetting was taken account of when fitting the data.

\section{Cluster Analysis}
Each of the 20 galaxy clusters and groups in the sample has a high quality
\emph{ROSAT} PSPC observation available. For several of the clusters, as
detailed in Table \ref{tab:sample}, \emph{ASCA} GIS data were also used.
The use of \emph{ASCA} GIS data is desirable for high temperature systems, as the
GIS has a bandpass that extends to much higher energies than the \emph{PSPC}.

Our cluster analysis works by fitting analytical models to the spectral
images from one or both of the instruments. The models parametrize either
the gas density and temperature, or the gas density and dark matter
density, as a function of radius. Dark matter as far as these models are
concerned is all gravitating matter apart from the X-ray emitting gas. Under
the assumption of hydrostatic equilibrium and spherical symmetry, the
equation
\begin{equation}
M(r)=-\frac{T(r)r}{G\mu}\left[\frac{dln\rho}{dlnr}+\frac{dlnT}{dlnr}\right]
\end{equation}
is satisfied \cite{fabricant84a}, and therefore the dark matter density
distribution can be calculated from the temperature distribution or vice
versa, if the gas density distribution is known. The models assume that the
systems are spherically symmetric and the dark matter models also assume
hydrostatic equilibrium. It is also assumed that the densities and
temperatures can be reasonably represented by analytical functions and that
the plasma is single phase (i.e. each volume element contains gas at just a
single temperature). The density in all the models is represented by a
core-index function of the form:
\begin{equation}
\rho(r)=\rho(0)\left[1+\left(\frac{r}{r_{c}}\right)^{2}\right]^{-\frac{3}{2}\beta}
\label{eq:beta}
\end{equation}
where $r_{c}$ is the core radius and $\beta$ is the density index. This has
been shown to be a good fit to observations of clusters \cite{jones84a}.
The temperature profile is parametrized using a linear function of the
form:
\begin{equation}
T(r)=T(0)-{\alpha}r
\label{eq:linear}
\end{equation}
where $\alpha$ is the temperature gradient. In the case of the dark matter
density parametrization we use a profile derived from numerical
simulations \cite{navarro95a} of the form:
\begin{equation}
\rho_{DM}(r)=\overline{\rho}_{DM}\left[x(1+x)^{2}\right]^{-1}
\label{eq:nfw}
\end{equation}
where $x=r/r_{s}$ and $r_{s}$ is a scale radius. Combining this with in gas
density distribution results in the total mass density distribution which
along with a temperature normalization parameter $T(0)$ allows the gas
temperature distribution to be calculated. The metallicity of the gas is
parametrized as a linear ramp in a similar way to the gas temperature.
The metallicity gradient was fixed at zero where only \emph{ROSAT} PSPC
data were used. The aim of using models that parameterize the gas
temperature both directly and indirectly, is to more fully explore the
parameter space available and so try to reduce the problem of implicit bias
associated with using a specific analytical model.

Our analysis also allows an optional extra cooling flow component to be
included in the models. This takes over from the normal density and
temperature parametrizations inside a cooling radius which is a fitted
parameter of the model. The density increases and the temperature decreases
as a powerlaw from the values at the cooling radius to the
\begin{table*}
\begin{minipage}{148mm}
\caption{Main parameters of the best fitting models for the sample. 
  The temperature and dark matter models have slightly different
  parameters. An asterisk in the CF column specifies that the model used a
  cooling flow component.}
\label{tab:models}
\begin{tabular}{lccccccccccc}
\hline
\hline
Cluster/Group&$\rho(0)$&$r_{c}$&$\beta$&T(0)&$\alpha$&$\rho_{DM}$(0)&$r_{s}$&CF\\
&(cm$^{-3}$)&(arcmin)&&(keV)&(keV arcmin$^{-1}$)&(amu cm$^{-3}$)&(arcmin)&\\
\hline
HCG 68 & 0.0161 & 0.28 & 0.44 & 0.86 & 0.036 & - & - & \\
HCG 97 & 0.119 & 0.04 & 0.41 & 1.05 & 0.020 & - & - & \\
HCG 62 & 0.138 & 0.03 & 0.36 & 1.49 & 0.023 & - & - & * \\
NGC 5044 Group& 0.009 & 1.66 & 0.49 & 1.21 & -0.005 & - & - & * \\
RX J0123.6+3315 & 0.121 & 0.10 & 0.43 & 1.50 & 0.022 & - & - & * \\
Abell 262 & 0.00725 & 1.45 & 0.39 & 1.45 & -0.081 & - & - & * \\
IV Zw 038 & 0.00116 & 2.77 & 0.38 & 2.39 & 0.043 & - & -  & \\
Abell 400 & 0.00189 & 3.83 & 0.51 & 1.68 & -0.014 & - & -  & \\
Abell 1060 & 0.00319 & 7.35 & 0.70 & 3.28 & - & 0.189 & 4.92 & * \\
MKW 3s & 0.0270 & 0.64 & 0.53 & 4.93 & 0.255 & - & - & \\
AWM 7 & 0.00418 & 5.53 & 0.60 & 2.88 & -0.094 & - & - & * \\
Abell 780 & 0.00855 & 1.69 & 0.67 & 4.05 & -0.203 & - & - & * \\
Abell 2199 & 0.00990 & 2.20 & 0.61 & 3.13 & -0.103 & - & - & \\
Abell 496 & 0.00504 & 3.24 & 0.64 & 6.34 & 0.140 & - & - & * \\
Abell 1795 & 0.0245 & 0.75 & 0.57 & 6.74 & - & 0.109 & 2.47 & \\
Abell 2218 & 0.00508 & 0.90 & 0.56 & 10.99 & 0.968 & - & - & \\
Abell 478 & 0.0236 & 0.84 & 0.62 & 8.318 & 0.445 & - & - & * \\
Abell 665 & 0.00754 & 0.73 & 0.52 & 13.76 & 1.465 & - & - & \\
Abell 1689 & 0.0290 & 0.60 & 0.73 & 12.31 & 0.002 & - & - & * \\
Abell 2163 & 0.00819 & 1.17 & 0.62 & 11.50 & 0.580 & - & - & \\
\hline
\end{tabular}
\end{minipage}
\end{table*}
centre, with fitted powerlaw indices. In the case of models that
parametrize dark matter density rather than temperature, no explicit
cooling flow temperature parameterization is needed, as the model permits
the derived temperature to drop at small radii. The density and temperature
powerlaws were flattened inside $r$ = 10 kpc, to prevent them going to
infinity.

The models described above, specify the density, temperature and
metallicity at each point in the cluster. Using the MEKAL hot coronal
plasma code \cite{mewe86a} it is then possible to compute the emission from
each volume element, and to integrate up the X-ray emission for each line
of sight through the cluster. This predicted emission is then convolved
with the response of the instrument in order to calculate the predicted
observation for the instrument. Standard energy responses and vignetting
functions were used for each instrument. Position and energy dependent
point spread functions were used. The \emph{ASCA} GIS PSF is obtained by
interpolating between several observations of Cyg X-1 at various positions
on the detector \cite{takahashi95a}. After folding the projected data
through the spatial and spectral response of the instrument and applying
vignetting, a predicted spectral image is obtained. This is then compared
with the observed spectral image, and the model parameters altered
iteratively, until a best fit is obtained. In cases where both \emph{ROSAT}
PSPC and \emph{ASCA} GIS data were used, the model was fitted to both
datasets simultaneously. This required careful adjustment to take account
of differences in the response and pointing accuracy of the different
telescopes. To achieve this the \emph{ASCA} GIS dataset was repositioned so
that the models fitted to same position as the \emph{ROSAT} PSPC. Our
analysis allows renormalization factors to be applied to the model
predictions to take account of gain variations between the different
instruments. A maximum likelihood method was used to compare the data with
the model predictions, as there are low numbers of counts in many bins of
the spectral image, and hence $\chi^{2}$ is inappropriate. Further details
of this cluster analysis technique can be found in \scite{eyles91a}.

Because of the large number of parameters in our models, typically $\geq
10$, the fit space for the models can be complicated.  It is necessary to
find the global minimum of the fit statistic in the fit space. Two
complementary methods were used to minimize the fit statistic and find the
best model fit. Initially a genetic algorithm \cite{holland75a} was used to
try to get close to the global minimum in the fit space. This works by
creating a population of solutions randomly distributed across the fit
space.  These solutions are then allowed to reproduce, by mutation
(altering parameters) or sexual reproduction (crossing over or averaging
parameters between parent solutions) with more chance of reproduction being
given to solutions giving better fits. Solutions with the poorest fits are
killed off as new solutions are created, and in this way the fitness of the
population improves through `natural selection'. This method is less likely
to get trapped in local minima in the fit space than conventional descent
methods. Once the locality of the global minimum is found a more
conventional modified Levenberg-Marquardt method \cite{bevington69a} was
used to find the exact position of the minimum in the fit space. By using
these two methods in conjunction the global minimum is much more likely to
be found.

Confidence intervals for the model parameters were calculated by perturbing
each parameter in turn from its best fit value, while allowing the other
fitted parameter to optimize, until the fit statistic increased by 1.  This
was done in the positive and negative directions for each fitted parameter,
to obtain the parameter offsets that correspond to this change in the fit
statistic. A change in the fit statistic of 1 corresponds to 1$\sigma$
confidence. All errors quoted below are 1$\sigma$.

The models used to derive the results presented below were the temperature
or dark matter model for each system that gave the best fit to the data.
Once the fitted models had been obtained it was possible to derive many
different system properties, including total gravitating mass and gas
entropy profiles. Throughout the following analysis we adopt $H_{0}$=50~km
s$^{-1}$ Mpc$^{-1}$ and $q_{0}$ = 0.5 , and show the $H_{0}$ dependence of
key results in terms of $h_{50}$ (=$H_{0}$/50). 

\begin{figure}
\centering{
\vbox{\psfig{figure=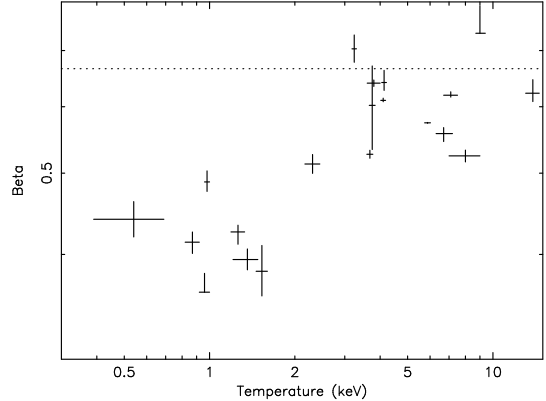}}\par
}
\caption{$\beta$ plotted against system temperature
  for 20 galaxy clusters and groups. The dotted line shows
  $\beta=\frac{2}{3}$.}
\label{fig:plot0}
\end{figure}

\section{Results}
The main parameters of the best fit model for each system in the sample are
shown in Table \ref{tab:models}. In this paper we will concentrate on the
departures from self-similarity in these systems and specifically the
entropy and energy of the intergalactic gas. A further paper is in
preparation which deals with other results from our sample. Before deriving
entropy profiles for the sample, the fitted parameters of the models
themselves were studied to see if they deviated from self-similar scaling
predictions.  The $\beta$ parameter in Equation \ref{eq:beta}, which is
essentially equivalent to the $\beta_{fit}$ parameter often used to fit
X-ray surface brightness profiles, showed a strong departure from
self-similarity in the low mass systems.  This is shown in Fig.
\ref{fig:plot0}. It can be seen that the gas density profiles of the
systems in the sample are not simply scaled versions of one another.  High
mass systems have $\beta$ values around the canonical value of
$\frac{2}{3}$. Low mass systems have significantly flatter gas density
profiles, with $\beta$ dropping to $\sim0.4$ for the galaxy groups which
agrees well with the \scite{helsdon99a} study of the surface brightness
profiles of galaxy groups. This is also supported by most recent studies of
galaxy clusters \cite{arnaud98a,jones99a} and is predicted by recent
simulations of energy injection into clusters
\cite{metzler97a,cavaliere98a}. However, \scite{mohr99a} fitted two
component core-index models to the surface brightness profiles of a sample
of galaxy clusters and found no dependence of $\beta_{fit}$ on temperature.
Our analysis also allows for the presence of a second central component
associated with a cooling flow, where necessary.  However the apparent
conflict between Fig. \ref{fig:plot0} and the results of \scite{mohr99a} is
resolved by the fact that their sample did not extend much below 3 keV, and
it can be seen from the figure that no significant trend above 3 keV is
seen in our sample. It should be noted that the $\beta$ values we derive
parametrize 3-dimensional gas density and are not directly comparable to
$beta$ values that parameterize X-ray surface brightness as isothermality
has not been assumed. In general the $beta$ values that we derive are
slightly lower than those derived from surface brightness profiles
\cite{mohr99a,ettori99a}. Some difference is to be expected as we do not
assume isothermality.

\begin{figure*}
\begin{minipage}{140mm}
\centering{
\vbox{\psfig{figure=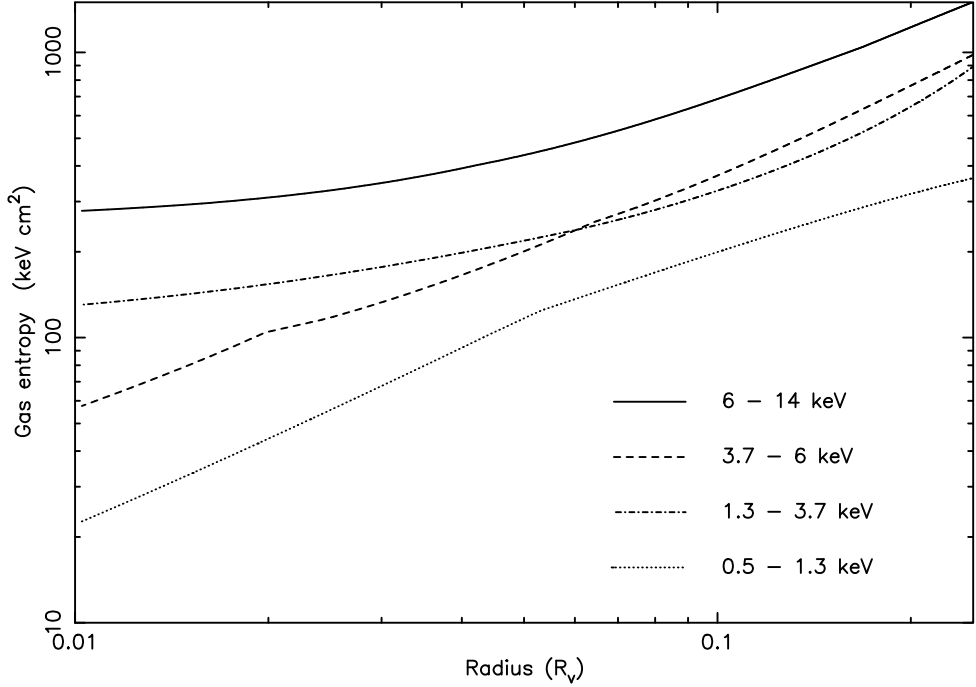}}\par
}
\caption{Mean gas entropy against radius scaled to the virial radius for the
  sample, grouped by system temperature. The solid line represents the five
  most massive systems (6-14 keV), through dashed (3.7-6 keV) and dash-dot
  (1.3-3.7 keV), to dotted (0.5-1.3 keV) for the lowest mass systems. The
  discontinuities in slope seen in some mean profiles, occur at the outer
  boundary of a central cooling flow component in the fitted model.}
\label{fig:plota}
\end{minipage}
\end{figure*}

\begin{figure*}
\begin{minipage}{140mm}
\centering{
\vbox{\psfig{figure=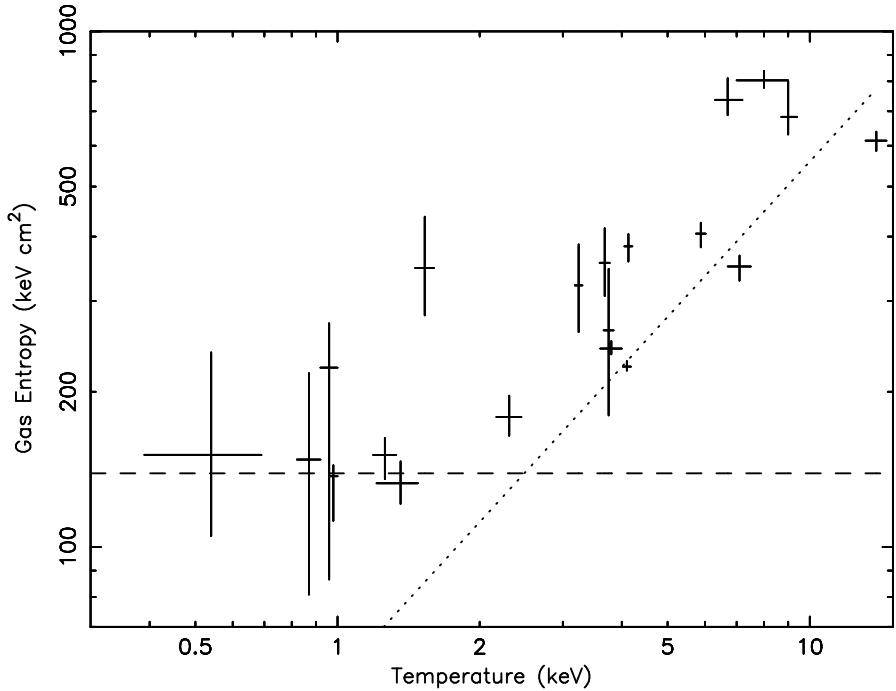}}\par
}
\caption{Gas entropy at 0.1R$_{v}$ against system temperature
  for 20 galaxy clusters and groups. The dotted line is a S$\propto$T fit
  to the systems above 4 keV excluding Abell 665 and Abell 2218 (see
  discussion in text). The dashed line is a constant entropy floor of
  139$\pm$7~keV~cm$^{2}$ fitted to the four lowest temperature systems.}
\label{fig:plot2a}
\end{minipage}
\end{figure*}

\begin{figure*}
\begin{minipage}{140mm}
\centering{
\vbox{\psfig{figure=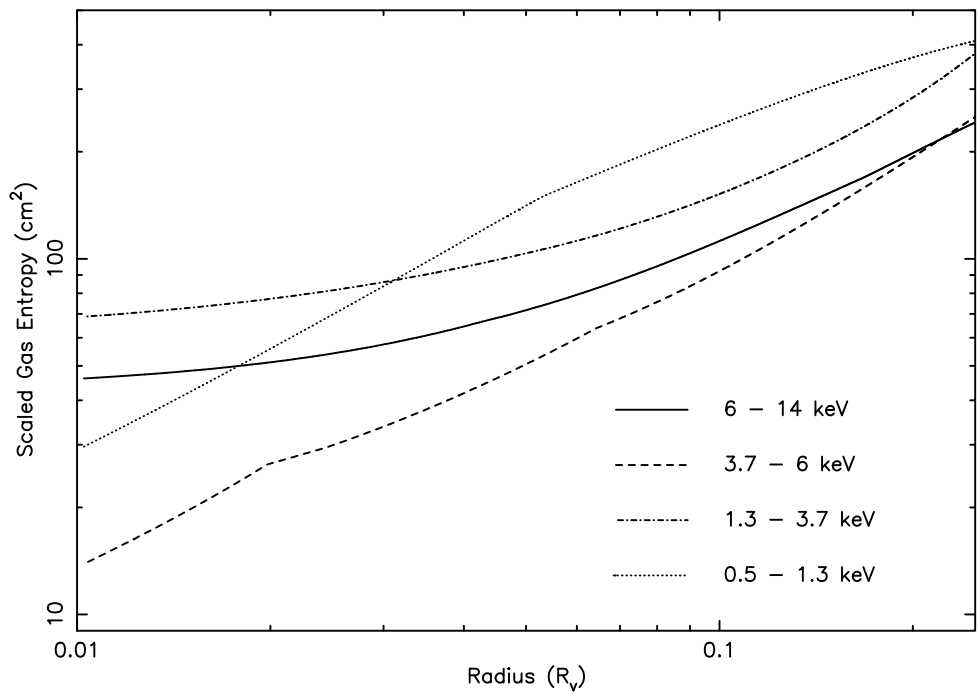}}\par
}
\caption{Mean gas entropy scaled by $T^{-1}(1+z)^{2}$ against radius scaled to
  the virial radius for the sample, grouped by system temperature. The solid
  line represents the five most massive systems (6-14 keV), through dashed
  (3.7-6 keV) and dash-dot (1.3-3.7 keV), to dotted (0.5-1.3 keV) for the
  lowest mass systems.}
\label{fig:plot1}
\end{minipage}
\end{figure*}

\begin{figure*}
\begin{minipage}{140mm}
\centering{
\vbox{\psfig{figure=plot2.ps}}\par
}
\caption{Scaled entropy at 0.1R$_{v}$ against system temperature
  for 20 galaxy clusters and groups. The dotted line shows the weighted
  mean scaled entropy of 54$\pm$3~cm$^{2}$ for the systems above 4
  keV excluding Abell 665 and Abell 2218 (see discussion in text).}
\label{fig:plot2}
\end{minipage}
\end{figure*}

\subsection{Excess entropy}

Gas entropy profiles as a function of radius were derived for the 20
systems. It is convenient to define `entropy' in terms of the observed
quantities, ignoring constants and logarithms, as 
\begin{equation}
S=\frac{T}{n_{e}^{2/3}}
\label{eq:entropy}
\end{equation}
where $T$ is the gas temperature in keV and $n_{e}$ is the gas electron
density. 
The radius axis of each profile was scaled to the virial radius of the
system, calculated using the formula
\begin{equation}
R_{v}=2.57\left(\frac{T}{5.1 \rm{keV}}\right)^{\frac{1}{2}}
(1+z)^{-\frac{3}{2}} \rm{Mpc}
\end{equation}
derived from numerical simulations \cite{navarro95a}. The profiles were
then grouped together by temperature and averaged in order to 
improve the clarity of
the figures and to high-light their temperature dependence. The mean
entropy profiles for groups of systems with similar temperatures are shown
in Fig.  \ref{fig:plota}. Each line in the figure is the average of the
profiles of 5 systems in a certain temperature range. It can be seen that
the most massive systems have the highest entropy gas. The gas entropy in
all systems shows a general increase with radius. This is to be expected,
as if the entropy declined with radius the gas would be convectively
unstable.  The profiles are similar to those seen in hydrodynamical
simulations such as those of \scite{metzler94a} and \scite{knight97a}. At
small radii the profiles are dominated by the effects of cooling flows in
many systems, resulting in a lowered central gas entropy.

To investigate the dependence of gas entropy on system temperature,
the entropy at 0.1~$R_{v}$ has been plotted against the mean system
temperature for each individual system. This is shown in Fig.
\ref{fig:plot2a}. A radius of 0.1 $R_{v}$ was chosen to be close to the
cluster centre (where shock heating is minimized), but to lie outside the
cooling region in all systems.  It can be seen that for the high
temperature systems the gas entropy is will appears to follow the expected
$S \propto T$ scaling. The dotted line is a powerlaw with a slope of unity
fitted to the systems with mean temperatures above 4~keV. It is clear that
the low temperature systems deviate from this trend and appear to flatten
out to a constant entropy floor. The dashed line is a constant gas entropy
fitted to the four lowest temperature systems and has a value of
139$\pm$7~${h_{50}}^{-\frac{1}{3}}$~keV~cm$^{2}$. This effect has
previously been noted by \scite{ponman98a} using isothermal assumptions.

In order to study the departures from self-similar scaling in more detail,
the profiles were scaled by a factor $T^{-1}(1+z)^{2}$, where $T$ is the
integrated system temperature and $z$ is the system redshift, and
overlayed. The $T^{-1}$ scaling should remove the effects of system mass,
as from Equation \ref{eq:entropy} it can be seen that `entropy' is directly
proportional to gas temperature. The $(1+z)^{2}$ scaling removes the effect
of the evolution of the mean density of the Universe, which has a
$(1+z)^{3}$ dependence, and results in systems that form at higher
redshifts being more dense.  This assumes that the systems formed at the
redshift of observation. The net result of this scaling is that the
profiles should fall on top of each other in the case of simple
self-similar scaling. The profiles were then grouped as before, resulting in
the mean profiles shown in Fig. \ref{fig:plot1}.

It can be seen from Fig. \ref{fig:plot1} that the scaled entropy profiles
of the sample do not coincide. In general the less massive systems have
higher scaled entropy profiles, with galaxy groups having the highest
scaled entropy values. This can be seen more clearly in Fig.
\ref{fig:plot2}, where the scaled entropy at 0.1 $R_{v}$ has been plotted
against the mean system temperature. The lower mass systems clearly show an
excess in scaled entropy over the high mass systems. In particular, systems
with temperatures above 4 keV appear to have a roughly constant scaled
entropy, while for systems with temperatures below 4 keV the scaled
entropy increases with decreasing temperature. A radius of 0.1 $R_{v}$ was
used as this lies outside the cooling flow regions of all the systems. 

Three of the systems stand out as being somewhat different from the general
trend. These are the clusters Abell 2218 and Abell 665, and the group IV Zw
038 (also known as the NGC 383 group). As well as lying above the trend in
Fig. \ref{fig:plot2}, they also show unusual scaled entropy profiles  
having the highest central scaled entropies in the sample.  Our fits
indicate that both of the clusters have very high temperature gradients, a
linear temperature fit (Equation \ref{eq:linear}) gives $\sim$6.2 keV
Mpc$^{-1}$ for Abell 665 and $\sim$4.3 keV Mpc$^{-1}$ for Abell 2218, which
are which make them very unusual compared to the rest of our sample.
Neither of these clusters has a significant cooling flow, and both Abell
2218 \cite{girardi97a} and Abell 665 \cite{markevitch96a} have been
suggested as being on-going or recent mergers. IV Zw 038 has a somewhat
lower temperature gradient of $\sim$1.5 keV Mpc$^{-1}$ although this is
still large, given the low mean temperature of this system.
\scite{komossa99a} have studied the X-ray emission of IV Zw 038 and
concluded that it is fairly relaxed. However \scite{sakai94a} studied
the distribution of galaxies around IV Zw 038 and concluded that the system
is highly substructured. It therefore appears that IV Zw 038 may also be an
ongoing or recent merger. As noted previously, transient flattening of
entropy profiles during mergers is seen in hydrodynamical simulations
\cite{metzler94a}.

\begin{figure}
\centering{
\vbox{\psfig{figure=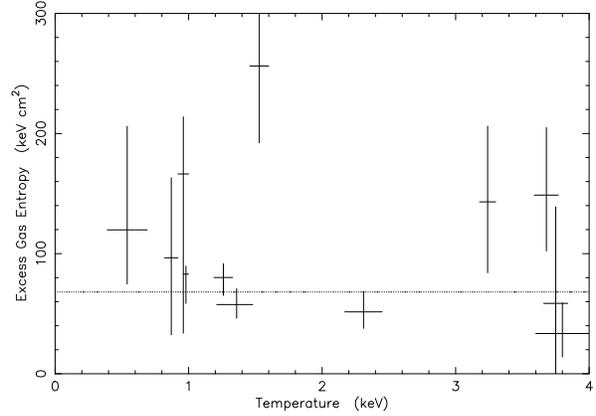}}\par
}
\caption{Excess entropy at $0.1 R_{v}$ against system temperature
for 12 systems with temperatures below 4 keV. The dotted line shows
the weighted mean value of 68$\pm$12 keV cm$^{2}$.}
\label{fig:plot3}
\end{figure}

\begin{figure}
\centering{
\vbox{\psfig{figure=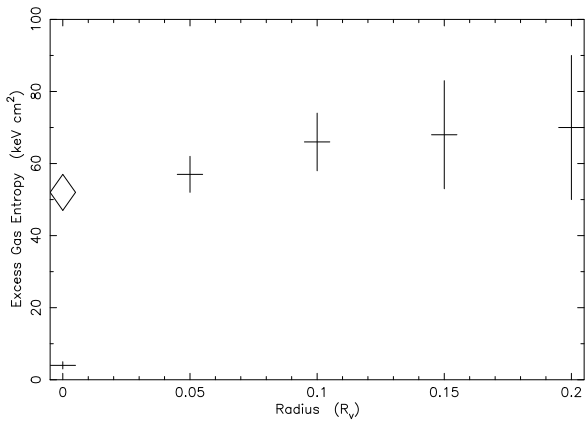}}\par
}
\caption{Mean excess gas entropy against radius for the 12 systems with
 mean temperatures below 4 keV. The diamond shows the central mean excess
entropy for the 3 systems without significant cooling.}
\label{fig:plot5}
\end{figure}

A mean value was calculated for the scaled entropy of the systems with
temperatures above 4 keV. The clusters Abell 2218 and Abell 665 were
excluded from this calculation due to their deviant behaviour. A weighted
mean value of 54$\pm$3 cm$^{2}$ was calculated. This was subtracted off the
entropies of the 12 systems with temperatures below 4 keV to calculate
their excess entropy. This (unscaled) excess entropy is plotted against
system temperature in Fig. \ref{fig:plot3}. The excess entropy shows no
trend with temperature and has a mean value of 68$\pm$12 keV cm$^{2}$. This
value drops slightly to 67 keV cm$^{2}$ if IV Zw 038 is excluded.
To investigate whether the excess gas entropy varies with radius this
procedure was repeated for radii from 0.0-0.2 $R_{v}$. It was not possible
to extend this analysis beyond 0.2 $R_{v}$ reliably, because the data for
the lowest mass systems does not extend that far due to their low surface
brightness. The variation of mean excess entropy against radius is shown in
Fig. \ref{fig:plot5}.

The mean excess entropy appears to be constant outside a central cooling
region which principally affects the innermost point in the figure. When
only the three systems without central cooling are plotted, the result is
the diamond in Fig. \ref{fig:plot5}.  This seems to confirm that the excess
entropy is distributed fairly evenly with radius and it is the effect of
cooling flows that causes the radial dependence seen in Fig.
\ref{fig:plot5}.  The cooling radii for these systems are $<$ 0.1 $R_{v}$
and it is to be expected that within cooling flows large amounts of entropy
will be lost as the gas cools.  The asymptotic value of excess entropy
outside the cooling region is $\sim$70 ${h_{50}}^{-\frac{1}{3}}$ keV
cm$^{2}$.

To investigate whether cooling flows are having any impact on the entropy
profiles of the systems at large radii (cf. discussion of the
\scite{knight97a} result in the introduction), excess entropy at 0.1
R$_{v}$ was compared to cooling flow size. It was possible to derive
reliable cooling flow mass deposition rates for 8 of the 12 systems with
temperatures below 4 keV (the remaining systems were consistent with no
cooling within errors or were not constrained by the analysis). This was
done using the equation,
\begin{equation}
\stackrel{.}{M}(i)=\frac{L(i)-[\Delta{h(i)}+\Delta{\phi(i)}]
\sum_{i'=1}^{i'=i-1}\stackrel{.}{M}(i')}
{h(i)+f(i)\Delta{\phi}}
\end{equation}
from \scite{white97a}, where $\stackrel{.}{M}(i)$ is the mass deposition
rate, $L(i)$ is the luminosity, $h(i)$ is the thermal energy per particle,
and $\phi(i)$ is the gravitational energy per particle in the radial bin
$i$. The $\Delta$ symbols represent a change in a quantity across a radial
bin. $f(i)$ is a factor that can be calculated to allow for the volume
averaged radius at which the mass drops out in the radial bin $i$. A value
of 1 was used for $f(i)$ for simplicity, which is consistent with previous
analysises \cite{white97a}. By integrating this equation out from the
centre of the system the mass deposition rate within any radius can be
calculated. The radius at which the cooling time equals the Hubble time,
1.3 $\times$ 10$^{10}$yrs for $H_{0}$ = 50 km s$^{-1}$ Mpc$^{-1}$, was used for
consistency with previous work. 

\begin{table}
\caption{Cooling flow mass deposition rates derived in this work compared
to rates from the literature. Asterisks indicate systems 
shown in Fig. \ref{fig:plot4}. The dashes in the second column indicate that
no reliable value could be derived from our analysis and the dashes in the
third column indicate that no value was available in the literature.}
\begin{minipage}{85mm}
\begin{center}
\begin{tabular}{lcc}
\hline
\hline
Cluster/Group&$\dot{M}$ this work&$\dot{M}$ literature\\
&($M_{\odot}yr^{-1}$)&($M_{\odot}yr^{-1}$)\\
\hline
HCG 68$\ast$&       $0.7^{+1.3}_{-0.5}$&      -\\         
HCG 97$\ast$&       $4^{+19}_{-3}$&           -\\           
HCG 62&       $6^{+95}_{-6}$&           $\sim$10\\
NGC 5044 Group$\ast$&$25^{+20}_{-3}$&         20-25\\       
RX J0123.6+3315$\ast$&$18^{+6}_{-3}$&         -\\            
Abell 262&    -&                        -\\                     
IV Zw 038&    -&                        $27^{+4}_{-3}$\\                
Abell 400$\ast$&         $7.2^{+0.5}_{-0.5}$&        $0^{+28}_{-0}$\\   
MKW 3s$\ast$&         $50^{+5}_{-5}$&                 $175^{+14}_{-46}$\\   
Abell 1060&        -&                           $15^{+5}_{-7}$\\                
AWM 7$\ast$&         $65^{+21}_{-21}$&           $41^{+6}_{-6}$\\            
Abell 780$\ast$&         $274^{+25}_{-24}$&          $264^{+81}_{-60}$\\    
Abell 496&         $105^{+17}_{-16}$&          $134^{+58}_{-85}$\\
Abell 2199&        $243^{+13}_{-10}$&          $154^{+18}_{-8}$\\
Abell 1795&        -&                           $321^{+166}_{-213}$\\
Abell 2218&        -&                           $66^{+76}_{-30}$\\
Abell 478&         $974^{+162}_{-132}$&        $616^{+63}_{-76}$\\
Abell 665&         -&                           $0^{+206}_{-0}$\\
Abell 1689&        $470^{+70}_{-180}$&         $0^{+398}_{-0}$\\
Abell 2163&        -&                           $0^{+256}_{-0}$\\
\hline
\end{tabular}
\end{center}
Note: The cooling flow mass deposition rates from the literature were
obtained from \scite{peres98a}, \scite{white97a}, \scite{david94a} and
\scite{ponman93a}.
\end{minipage}
\label{tab:cooling}
\end{table}

\begin{figure}
\centering{
\vbox{\psfig{figure=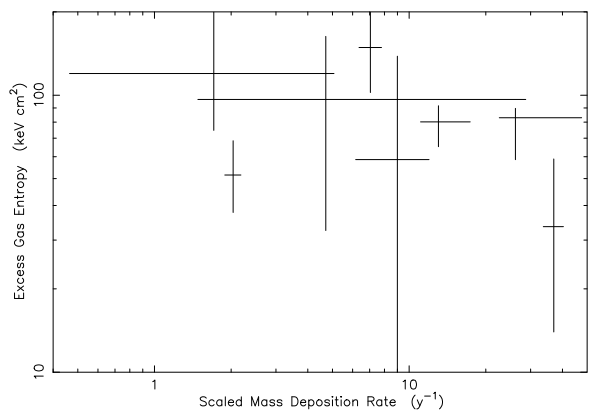}}\par
}
\caption{Excess entropy at $0.1 R_{v}$ against cooling flow mass
deposition rate scaled by $T^{-3/2}$
for 8 systems with temperatures below 4 keV.}
\label{fig:plot4}
\end{figure}

The cooling flow mass deposition rates derived from our analysis for the
whole sample are listed in Table \ref{tab:cooling} along with values taken
from the literature. In general there is good agreement between the values
we derive and previously derived values. The cooling flow mass deposition
rates for the 8 systems with temperatures below 4 keV were then scaled by
$T^{-3/2}$, which is proportional to $M^{-1}$, to scale the cooling flows
to the system size. The scaled mass deposition rates are therefore
proportional to the fraction of the cluster mass that is cooling out per
year. These scaled mass deposition rates have been plotted against excess
entropy in Fig.  \ref{fig:plot4}. It can be seen that there is no
appreciable correlation between excess entropy and scaled mass deposition
over more than an order of magnitude range in scaled mass deposition rate.
The weighted mean value for the excess entropy in the systems with no
measurable cooling is 67 $\pm$ 15 keV cm$^{2}$, almost identical to the
mean for all the systems with temperatures below 4 keV. These results
confirm that cooling is not driving the trend seen in Fig. \ref{fig:plot2}.

\subsection{Excess energy}

\scite{ponman98a} assumed that their systems were isothermal, and so
were unable to measure the extra energy in the IGM which gives rise to
this excess entropy. Their analysis was therefore based on a rough
estimate of the likely energy injection based on the {\it assumption}
that it was caused by supernova-driven galactic winds. Here,
because of our spatially resolved temperature profiles, we can actually
attempt to {\it measure} the injected energy and then compare it to the energy
expected from galactic winds or other heating mechanisms.

\begin{figure*}
\begin{minipage}{140mm}
\centering{
\vbox{\psfig{figure=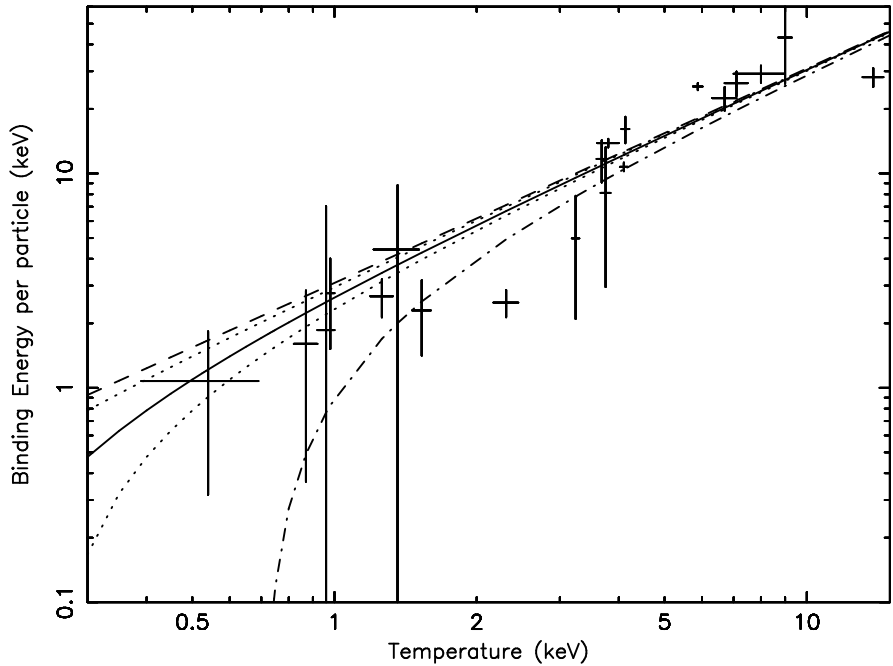}}\par
}
\caption{Mean binding energy per particle of the central 0.004$M_{200}$
  of gas against system temperature for 20 galaxy clusters and groups. The
  dashed line shows a $E \propto T$ line fitted through the systems with
  mean temperatures greater than 4 keV. The solid line shows our best
  estimate of the $E=AT-\Delta{E}$ relationship (for $\Delta{E}=0.44$~keV)
  with 1$\sigma$ errors (dotted).The dot-dashed line is the result
  obtained when the unreliable point for the cluster Abell 400 is included 
  in the fit (see discussion in text).}
\label{fig:plot6}
\end{minipage}
\end{figure*}

The excess energy is composed of two parts: extra thermal energy, and
reduced gravitational binding energy. Due to the fact that our data do
not extend beyond 0.2 $R_{v}$ for the lowest mass systems, it was not
possible to calculate the total binding energy of the gas within the
virial radius for the entire sample, and since energy injection will
change the gas distribution, considering the binding energy of the gas
within a fixed {\it fraction} of the virial radius will be misleading.
Instead, we investigate the binding energy of gas constituting a fixed
fraction of the virial mass of each system. If the gas distributions
of the systems were self-similar, this would translate into gas within
a fixed fraction of the virial radius, but it can be seen from Fig.
\ref{fig:plot0} the gas distributions of the systems are not
self-similar. In order to calculate the virial masses of the systems
from their mean temperatures the formula
\begin{equation}
M_{200}=10^{15}\left(\frac{T}{5.1 \rm{keV}}\right)^{\frac{3}{2}}
(1+z)^{-\frac{3}{2}}\rm{M}_{\odot}
\label{eq:mass}
\end{equation}
was used \cite{navarro95a}, which is derived from numerical simulations. A
fixed fractional gas mass of 0.004 $M_{200}$ was used, which was found to
correspond to a fraction of the virial radius between 0.064 and 0.226 for
the systems in the sample. 

The mean binding energy per particle of the central 0.004 $M_{200}$ of gas
for the sample is plotted against temperature in Fig. \ref{fig:plot6}. If
the systems were self-similar then the binding energy per particle would be
directly proportional to the temperature, and this relation, fitted to
systems with mean temperatures greater than 4 keV, is shown by the dashed
line. The uniform injection of a constant amount of excess energy per unit
system mass will result in a relation of the form
\begin{equation}
E = AT-\Delta{E}
\label{eq:phi}
\end{equation}
where $E$ is the binding energy per particle, $T$ is the mean gas
temperature, $\Delta{E}$ is the injected energy per particle and A is a
constant.

Using the function in Equation \ref{eq:phi} results in a best fit value
$\Delta E$ = 2.2~keV per particle, shown in Fig. \ref{fig:plot6} as a
dot-dash line. This result is clearly unreasonably large, as it would
preclude the presence of significant hot gas in systems with virial
temperatures less than $\sim$ 1.5 keV. It can be seen from Fig.
\ref{fig:plot6} that this model line underestimates the observed binding
energy in almost all the cooler systems. This result is being driven by one
system, Abell 400, which has an exceptionally small gas binding energy,
with a small statistical error.  However, \scite{beers92a} have studied the
galaxy distribution in Abell 400 in detail, and concluded that it is highly
subclustered, with two major subclusters essentially superposed on the
plane of the sky. Hence the apparently relaxed X-ray morphology in this
system is probably misleading, and our derived energy and entropy values
for the cluster are unsafe.

Excluding Abell 400 from our analysis, gives a much lower value for the
fitted value of excess energy: $\Delta{E}$ = 0.44 keV per particle,
corresponding to a preheating temperature of $T = 0.3 \pm 0.2$~keV. The fit
is shown as a solid line in Fig. \ref{fig:plot6}, along with a formal
1$\sigma$ confidence interval. Clearly this estimate of the excess energy
is subject to large statistical and systematic errors at present, and a more
accurate result should be available in due course from studies with the new
generation of X-ray observatories. However, as we will discuss below, a
value of $\sim 0.4$~keV per particle agrees well with recently developed
preheating models, and with estimates based on the metallicity of the IGM.

\begin{figure}
\centering{
\vbox{\psfig{figure=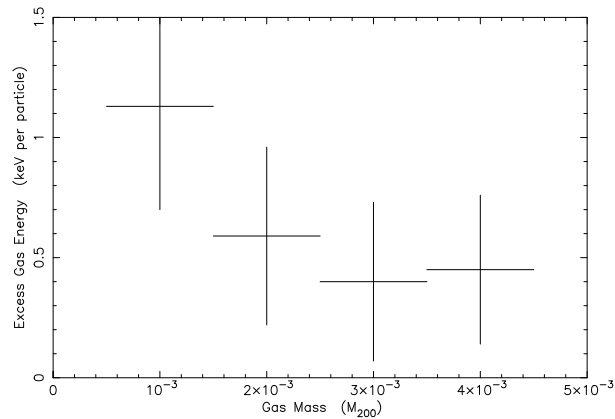}}\par
}
\caption{Excess gas energy derived for a range of fractional gas masses.
Since the gas mass is integrated from the cluster centre, the points
in the plot are not statistically independent.}
\label{fig:plot7}
\end{figure}

To investigate whether this measured injection energy shows any radial
dependence, the above procedure was repeated for a number of different
fractional gas masses. The results are plotted in Fig. \ref{fig:plot7}. At
small radii, the measured excess energy in the gas is affected by the
presence of cooling flows, which effectively scales up the whole of the
right hand side of Equation \ref{eq:phi} due to the increased central
concentration of the gas, resulting in a higher inferred value for
$\Delta{E}$. However, it can be seen that the effects of this distortion
are confined to $M_{\rm gas}<0.003 M_{200}$, and that the asymptotic value
of excess energy outside the cooling region is $\sim$ 0.4 keV.
Extrapolation of the models to larger fractional gas masses is highly
uncertain and would result in large systematic errors as it would encompass
gas well beyond the data in the low mass systems.

Since $P d\!V$ work and shock heating can move energy around within the
IGM, the excess energy per particle evaluated within a subset of the total
gas mass will not necessarily equal the value which would be obtained if we
could extend our analysis to cover the whole of the intracluster medium. A
simple model involving a flattened $\beta$-model gas distribution in
hydrostatic equilibrium within a NFW (Equation \ref{eq:nfw}) potential,
suggests that our result derived from the innermost 0.004$M_{200}$ of the
gas, may {\it overestimate} the excess energy, integrated over the ICM, by
a factor of $\sim$2.

Our analysis assumes that Equation \ref{eq:mass} holds even in the lowest
mass systems. Semi-analytical models of the effects of preheating, by
\scite{balogh98a} and \scite{cavaliere98a} indicate that preheating has
little effect on gas temperature except in systems with virial temperatures
close to the preheating temperature. The mass-temperature relations in both
the \scite{balogh98a} and \scite{cavaliere98a} studies deviate
significantly from the expected $M\propto T^{3/2}$ only at $T<0.8$~keV.
Only one member of our sample, HCG\,68, with a mean gas temperature of 0.54
keV, lies in this region.  To investigate the possibility that this point
in Fig. \ref{fig:plot6} may have been significantly affected, we derived the
mass of this system from our fitted model. Due to fact that the data extend
to only $\sim$ 0.2 R$_{v}$, this involves considerable extrapolation out to
the virial radius, with an associated (and uncertain) systematic error. The
mass derived from our fitted model was $2.4\times 10^{13}$~M$_{\odot}$,
compared to a value of $3.45\times 10^{13}$~M$_{\odot}$ from Equation
\ref{eq:mass} using the mean temperature of the system. If we have
overestimated $M_{200}$ for this system, then the gas mass we have
considered will be too large, and its binding energy (which decreases with
radius) will be too low.  Using $M_{200}=2.36 \times 10^{13}$M$_{\odot}$
instead, would result in the derived binding energy of the 0.004$M_{200}$
of gas being increased by 13\%. This systematic error is much less than the
statistical error on the point and so should have a minimal effect on the
fit. Any effect would be in the direction of reducing the injection energy.

The excess energy we have derived can be compared to what might reasonably
be available from galaxy winds. Assuming that the galaxy wind ejecta have
approximately solar metallicity, it appears that this gas has been diluted
by a factor of $\sim$3-5 with primordial gas, to arrive at the typical
metallicities of 0.2-0.3~solar, seen in galaxy groups and clusters
\cite{fukazawa98a,finoguenov99a}. A final excess of $\sim$0.4~keV per
particle after dilution, therefore implies an injected wind velocity of
$\sim 1000$~km~s$^{-1}$, assuming that the energy of the injected gas is
dominated by its bulk flow energy. Studies of local ultraluminous starburst
galaxies show outflows of cool emission line gas with velocities of a few
hundred km~s$^{-1}$, and models suggest terminal velocities for the hot gas
of a few thousand km~s$^{-1}$ \cite{heckman90a,suchkov94a,tenorio97a}.
Galactic winds therefore seem capable of providing the energy we observe.

\subsection{Constraints on preheating}

As both the excess entropy and preheating temperature of the ICM have been
measured, it can be seen from the definition of entropy in Equation
\ref{eq:entropy}, that it should be possible to derive the electron density
$n_{e}$ at which the energy was injected.  The details of the energy
injection process itself do not matter, provided that sufficient mixing of
the gas has subsequently occurred to distribute the energy uniformly at the
time of observation.  The inferred injection density is
\begin{equation} 
n_{e}=\left(\frac{\Delta{T}}{\Delta{S}}\right)^{\frac{3}{2}}
\label{eq:ne}
\end{equation}
where $\Delta{T}$ and $\Delta{S}$ are the changes in gas temperature
and entropy. Using the values obtained above for these quantities,
we derive an electron density at the time of injection of
$3\times 10^{-4}{h_{50}}^{\frac{1}{2}}$cm$^{-3}$. This is about an
order of magnitude lower than the mean gas density in cores of systems
without cooling flows, suggesting that the energy must have been
injected before these systems were fully formed.

However, if the entropy injection took place before the systems collapsed
it may have affected the shock heating efficiency in the low mass systems,
reducing the amount of entropy the shocks produced. In the extreme case,
shock heating could have been totally suppressed in the lowest mass
systems, in which case they would have collapsed adiabatically and their
present entropy would essentially be the total injected entropy. The degree
to which shocks have increased the entropy in the lowest mass system is not
at all clear.  However it should be noted that even in the lowest mass
systems in Fig.  \ref{fig:plot2}, the gas entropy is rising with radius
outside the cooling region, suggesting that some shock heating has taken
place.  The resolution of this problem will require detailed hydrodynamic
simulations of the formation of galaxy groups which is not available at
present. We therefore consider our previous result to be a lower bound on
the excess entropy in these systems and the measured entropy floor ($\sim$
140 keV cm$^{2}$) in Fig. \ref{fig:plot2a} to be an upper bound, applying
in the case where shock heating is totally suppressed.  For this second
case, Equation \ref{eq:ne} results in an even lower value of $1\times
10^{-4}{h_{50}}^{\frac{1}{2}}$cm$^{-3}$ for the density at which the
entropy is injected.

Even if the injection took place outside a collapsed system, it must have
occurred after the mean density of the Universe dropped to 1-3$\times
10^{-4}$cm$^{-3}$, as before this, even uniformly distributed gas would be
too dense to produce the measured entropy change from the available energy.
Using the value for the baryon density of the Universe derived from Big
Bang nucleosynthesis, $\Omega_{b}h_{50}^{2}=0.076\pm0.0096$
\cite{burles99a}, and the fact that the density of the Universe scales as
$(1+z)^3$, it follows that the mean electron density of the Universe would
be less than $3\times 10^{-4}$cm$^{-3}$ when $z<10$ and less than $1\times
10^{-4}$cm$^{-3}$ when $z<7$.

Hence we conclude that the entropy injection must have taken place after $z
\sim 7-10$, depending on the assumed amount of shock heating in low mass
systems, but before the galaxy systems have fully formed. In fact it is
likely that the baryons in these systems have always been in overdense
regions of the Universe, and therefore the entropy injection probably took
place at a considerably lower redshift than this conservative upper limit.
If our value of 0.44 keV per particle for the excess energy is an
overestimate, as discussed in Section 5.2, this would have the effect of
lowering the inferred gas density at injection, and reducing our redshift
limit.

This all assumes that the gas cannot expand as the energy is injected, i.e
an isodensity assumption. This will be true if the energy injection takes
place at high redshift when the density field of the Universe is still
fairly smooth and there is effectively nowhere for the gas to expand to.
However if the energy injection takes place at lower redshift in partially
formed systems, the gas may expand in the potential of the system. A more
realistic scenario in this case is that the energy is injected under
constant pressure, i.e.  it is isobaric. In the isobaric case the resulting
entropy change will be higher than the isodensity case, since density drops
as the injection proceeds.

To quantify the possible error involved in assuming that the gas does not
expand as the energy is injected, we investigate the difference in entropy
change between the case of isodensity and isobaric energy injection. The
entropy changes for the two cases will be:
\begin{enumerate}
\item{Isodensity - As the density does not change the only effect on the
entropy will be due to the change in temperature of the gas. The entropy
change ${\Delta}S$ will therefore be:
\begin{equation}
{\Delta}S=\frac{{\Delta}T}{n_{e}^{2/3}}
\end{equation}
where ${\Delta}T$ is the change in temperature. If the gas cannot expand
the temperature change will be related to the injected energy by the
equation
\begin{equation}
{\Delta}T=\frac{2}{3}{\Delta}E
\label{eq:energy}
\end{equation}
where ${\Delta}E$ is the injected energy per particle. The entropy change
for a given injected energy will therefore be
\begin{equation}
{\Delta}S=\frac{2}{3}\frac{{\Delta}E}{n_{e}^{2/3}}
\end{equation}
} 
\item{Isobaric - As the pressure remains constant the equation:
\begin{equation}
n_{0}T_{0}=n_{1}T_{1}
\end{equation}
will be satisfied, where $n_{0}$ and $n_{1}$ are the initial and final
electron densities and $T_{0}$ and $T_{1}$ are the initial and final temperatures
and so using the definition of entropy in Equation \ref{eq:entropy} the
change in entropy in terms of the change in temperature will be:
\begin{equation}
{\Delta}S=\frac{{\Delta}T}{n_{e}^{2/3}}\frac{(\gamma^{5/3}-1)}{(\gamma-1)}
\end{equation}
where $\gamma=\frac{T_{1}}{T_{0}}$ and $n_{e}=n_{0}$, the initial density.
However as work is done expanding the gas, the temperature change will not
be related to the injected energy as in Equation \ref{eq:energy} but will
be
\begin{equation}
{\Delta}T=\frac{2}{5}{\Delta}E,
\end{equation}
and so the entropy change for a given injected energy is
\begin{equation}
{\Delta}S=\frac{2}{5}\frac{{\Delta}E}{n_{e}^{2/3}}\frac{(\gamma^{5/3}-1)}{(\gamma-1)}.
\end{equation}
}
\end{enumerate}
The ratio of the changes in entropy between the isodensity and isobaric
case, for a given injected energy, is therefore:
\begin{equation}
\frac{{\Delta}S_{isobar}}{{\Delta}S_{isoden}}=\frac{3}{5}
\frac{(\gamma^{5/3}-1)}{(\gamma-1)}
\label{eq:ratio}
\end{equation}
and depends only on the value of $\gamma$, the ratio of the final to initial
temperatures. This ratio, given by Equation \ref{eq:ratio}, is shown in
Table \ref{tab:ratio} for a range of values of $\gamma$.
\begin{table}
\caption{The ratio of entropy changes in the isodensity and isobaric cases
  as a function of the ratio of final to initial temperatures, $\gamma$.}
\begin{center}
\begin{tabular}{lc}
\hline
\hline
$\gamma$&$\frac{{\Delta}S_{isobar}}{{\Delta}S_{isoden}}$\\
\hline
2&1.3\\
5&2.0\\
10&3.0\\
100&13.1\\
1000&60.1\\
\hline
\end{tabular}
\end{center}
\label{tab:ratio}
\end{table}

At high redshift, where the initial temperature of the gas is low, the
value of $\gamma$ will be large. However at high redshift the density field
should be fairly smooth and so the isodensity assumption should be a fairly
good one. At lower redshift, where the isobaric case will be more
realistic, the initial temperature of the gas in these partly formed
systems will be similar to the temperature change ($\sim$ 0.3 keV)
resulting from the entropy injection, and so $\gamma$ will be close to
unity. It can be seen from Table \ref{tab:ratio} that when $\gamma$ is
close to unity the difference between the isodensity and isobaric case is
small and so the isodensity result should still be a reasonable
approximation.  We conclude that the entropy increase should only be
slightly underestimated by the isodensity analysis given above, and hence
that the density limit of $3\times 10^{-4}$cm$^{-3}$ cannot be pushed
significantly higher by allowing for expansion of the gas during
preheating.

\section{Discussion}
It is clear from Figures \ref{fig:plot1} and \ref{fig:plot2} that systems
with integrated temperatures below 4 keV show signs of having excess
entropy in their intracluster gas over what would be expected from the
simple self-similar model. It can further be seen from Fig. \ref{fig:plot3}
that the amount of excess entropy does not depend systematically on the
system temperature and, from Fig.  \ref{fig:plot5}, it has an approximately
constant value outside the central cooling flow regions.  The average
excess entropy outside the cooling flow region lies in the range
70-140${h_{50}}^{-\frac{1}{3}}$ keV cm$^{2}$. The upper limit, where shocks
are totally suppressed in low mass systems, is comparable with the result of
\scite{ponman98a} who obtained a value of 100${h_{100}}^{-\frac{1}{3}}$
(126${h_{50}}^{-\frac{1}{3}}$)~keV~cm$^{2}$ for the assumption of total
shock suppression. This new upper limit on the entropy should be more
reliable as it does not rely on the assumption of isothermality of the
intracluster gas that \scite{ponman98a} had to use. Our analysis also
sets a lower bound on the entropy for the case where the shock heating
is not affected.

It is also interesting to compare our measured value for the excess entropy
against the value assumed in various theoretical models of entropy
injection in galaxy systems. For instance \scite{balogh98a} assume a
constant entropy injection value of $\sim $350 keV cm$^{2}$ in order to
reproduce the steepening in the $L$-$T$ relation for galaxy groups.
\scite{tozzi99a} argue that to steepen $L$-$T$ at $\sim$ 0.5-2 keV, entropy
injection in the range 190-960 keV cm$^{2}$ is needed. Both these values
are somewhat higher than our measured range, but considering the 
simplified nature of these models, the similarity is encouraging. It will
be interesting to see whether more sophisticated models can match the group
$L$-$T$ relation using the lower values of entropy we observe.

A number of models work on the assumption of some specific amount of energy
injection into the gas. These can be compared with the amount of
excess energy we observe to be present in galaxy systems.
\scite{cavaliere97a} and \scite{cavaliere98a} assume that the gas in galaxy
systems is preheated to a temperature of 0.5 keV which is comparable to
our measured value. \scite{wu99a} obtain energy input of $<0.1$~keV
per particle from SN heating within most of their
hierarchical merger model runs. However, it is not clear that this
represents a hard limit, since these authors assumed that gas can only
be heated to the escape velocity of their galaxy halos.
\scite{wu98a,wu99a} also find that 
an injected
energy per particle of $\sim$1-2 keV is required to reproduce the slope
of the cluster $L$-$T$ relation \cite{david93a}. This may indicate that
the preheating required to match the steepening in $L$-$T$ at $T<1$~keV
does not provide a solution to the departure of the cluster relation
from the self-similar result, $L\propto T^2$. For example, it is clear
that the model of \scite{cavaliere98a}, which provides a good match to
the group data, fails to reproduce the slope of the $L$-$T$ relation at
high temperatures (see their Fig.9). Additional effects may be at work
-- for example \scite{allen98a} have demonstrated that allowing for the
impact of cooling flows flattens the $L$-$T$ relation for rich clusters
towards $L\propto T^2$.

The floor entropy of 70-140 ${h_{50}}^{-\frac{1}{3}}$~keV is small compared
to the entropy of the 8 systems with temperatures of 4 keV or above, which
averages 380 keV cm$^{2}$ (at 0.1 $R_{v}$). Hence our results are
consistent with the idea that an approximately constant amount of excess
entropy, $\sim$70-140${h_{50}}^{-\frac{1}{3}}$ keV cm$^{2}$, is present in
{\it all} of the systems, but is only noticeable in systems where it
constitutes a large fraction of the total entropy, i.e. in systems with
temperatures below 4 keV. From Fig. \ref{fig:plot5} it can be seen that
there is little evidence for any dependence of the excess entropy on radius
outside the central cooling region.  This suggests that the process
involved in injecting entropy into the systems does so fairly uniformly, at
least within $0.25 R_{v}$.

From Fig. \ref{fig:plot6}, the gas in low mass systems is significantly
less tightly bound than would be expected from self-similar scaling.
Combining the excess entropy and energy requirements leads us to conclude
that the energy was injected at $z<$7-10, but before cluster collapse.
Possible candidates for the source of this extra energy are preheating by
quasars, population III stars or galaxy winds. It is known that since
recombination at z $\sim 1400$, the intergalactic medium has been
re-ionized. This re-ionization is normally assumed to be caused by quasars
or an early epoch of star formation. However analytical models of these
processes \cite{valageas99a,tegmark93a} suggest that the IGM will only be
heated to $\sim$ $10^{4}$-$10^{5}$K, resulting in an entropy change that is at
least an order of magnitude lower than the measured value. In contrast,
energy injected by supernovae associated with the formation of the bulk of
galactic stars should be much more significant \cite{white91a,david91a}.

The likely energies involved can be estimated from observed metal
abundances in the intracluster gas. The major uncertainty here lies in
establishing the contributions from supernovae of type Ia and type II,
which have very different ratios of iron yield to energy \cite{renzini93a}.
Recent studies with \emph{ASCA} \cite{finoguenov99a,finoguenov99b} in which
contributions from SNIa and SNII have been mapped in a sample of groups and
clusters, by tracing the abundance of iron and alpha elements, leads to the
conclusion that SNIa provide a significant contribution to the iron
abundance, particularly in galaxy groups. The supernova energy associated
with the observed metal masses by \scite{finoguenov99a} are in good
agreement with the energy of $\sim 0.4$~keV per particle derived above on
the basis of the observed energy excesses. This is also similar to the
preheating involved in the models of \scite{cavaliere97a},
\scite{cavaliere98a} and \scite{balogh98a} supporting the idea that the
similarity breaking we see in the intracluster gas does result from
preheating associated with galaxy formation.

With the forthcoming availability of data from \emph{Chandra} and
\emph{XMM}, much more detailed studies of the abundance and entropy
distributions of galaxy systems will become possible. This will allow
deviations from mean trends to be studied in detail. Since galaxy winds
will inject both energy and metals, whereas processes such as ram pressure
stripping will lead to metal enrichment without heating, studies with these
new X-ray observatories should throw a great deal of light on the
evolutionary history of galactic systems and the galaxies they contain.

\section*{Acknowledgments}
We thank Peter Bourner for his contribution to the preliminary data
analysis, and the referee for a number of useful suggestions.  Discussions
with Richard Bower, Mike Balogh, Alfonso Cavaliere and Kelvin Wu have
helped to clarify the relationship between the observations and preheating
models. This work made use of the Starlink facilities at Birmingham, the
LEDAS database at Leicester and the HEASARC database at the Goddard Space
Flight Centre. EJLD acknowledges the receipt of a PPARC studentship.

\bibliography{entropy}

\label{lastpage}

\end{document}